\documentclass[11pt,superscriptaddress,aps,prd,preprint]{revtex4}

\everymath{\displaystyle}
\usepackage{graphicx}

\usepackage{amsmath,amssymb,mathrsfs}

\newcommand{\bea}{\begin{eqnarray}}
\newcommand{\eea}{\end{eqnarray}}

\begin{document}

\title{On Gravitational Stefan-Boltzmann Law and Casimir Effect in FRW Universe}

\author{A. F. Santos}\email[]{alesandroferreira@fisica.ufmt.br}
\affiliation{Instituto de F\'{\i}sica, Universidade Federal de Mato Grosso,\\
78060-900, Cuiab\'{a}, Mato Grosso, Brazil}

\author{S. C. Ulhoa}\email[]{sc.ulhoa@gmail.com}
\affiliation{International Center of Physics, Instituto de F\'isica, Universidade de Bras\'ilia, 70910-900, Bras\'ilia, DF, Brazil} \affiliation{Canadian Quantum Research Center,\\
204-3002 32 Ave Vernon, BC V1T 2L7  Canada}

\author{E. P. Spaniol}
\email{spaniol.ep@gmail.com} \affiliation{UDF Centro Universit\'ario
and Centro Universit\'ario de Bras\'ilia UniCEUB, Bras\'ilia, DF,
Brazil.}

\author{Faqir C. Khanna\footnote{Professor Emeritus - Physics Department, Theoretical Physics Institute, University of Alberta\\
Edmonton, Alberta, Canada}}\email[]{khannaf@uvic.ca}
\affiliation{Department of Physics and Astronomy, University of Victoria,\\
3800 Finnerty Road Victoria, BC, Canada}

\begin{abstract}

Both Stefan-Boltzmann law and the Casimir effect, in a universe described by the FRW metric with zero curvature, are calculated. These effects are described by Thermo Field Dynamics (TFD). The gravitational energy-momentum tensor is defined in the context of Teleparallel Equivalent to General Relativity (TEGR).
Each of the two effects gives a consistent prediction with what is observed on a cosmological scale. One of the effect establishes a minimum range for the deceleration parameter. While another leads to the conclusion that a possible cosmological constant has a very small order of magnitude.

\end{abstract}
\maketitle

\date{\today}
\section{Introduction} \label{sec.1}

The introduction of temperature in the
gravitational field has been successfully implemented recently 
\cite{gravTFD}. Thermo Field Dynamics (TFD) was used for this purpose which is an
approach that allows both a temporal evolution of the field at finite temperature.
It is an advantage over the historical approach that associates time with temperature
\cite{matsubara}.  A gravitational field at finite temperature is a theory of quantum gravity since TFD uses creation and
annihilation operators. The field propagator is the fundamental entity of the thermalization
process.  It is interesting to note that such an approach associates
the temperature with space such that a universe with zero temperature
will not be expected \cite{flatTFD}.  Absolute zero temperature will not be
natural even in flat space.  Within the scope of TFD, there is a
topological structure that allows treating effects such as
diverse as the Stefan-Boltzmann law and the Casimir effect on an
equal footing. An area of intense investigation into the implications of various aspects of quantum gravity is black hole thermodynamics.  Whether in the investigation of entropy of black holes, or in the understanding of the information paradox.  It was recently investigated how the evaporation process of a black hole generates an entanglement between quantum fields and geometry, this yields a modified Page curve that can have implications for several theories of quantum gravity \cite{4}.  It has also been shown that the structure of TFD plays a key role in this approach \cite{5}. The TFD appears to be a promising theory of quantum gravity.

It is necessary to thermalize the energy-momentum tensor of the field in addition to a propagator. The standard
model of gravitation is problematic.  In the
construction of gravity at finite temperature, an alternative theory
of gravitation is used, Teleparallelism Equivalent to General
Relativity (TEGR) \cite{maluf}.  In TEGR the problem of
gravitational energy is well established, as well as other conserved
quantities.  As a result, gravitational entropy is introduced as a
direct consequence of Maxwell's relationships involving
gravitational pressure.  Normally this gravitational entropy may be
seen as a fundamental quantity when made equal to Hawking's
expression induces a temperature of the black hole event horizon
different from that commonly accepted. The whole space-time
has a finite temperature, not just the event horizon of a black
hole.  Then there is a smooth transition from singularity to
infinity \cite{entropy}. We must note that TEGR is a formulation of gravitation that takes into account local Lorentz's symmetry, such a dependence appears in the field equations that are entirely equivalent to Einstein's equations.  On the other hand, recently, proposals have emerged that attribute the local Lorentz symmetry to the spin connection \cite{7,8,9}, this line of investigation has received some criticism and in our opinion still requires further investigation \cite{10}.  In TEGR the conserved quantities are sensitive to the global Lorentz transformations and that is the limit of our approach.

One of the major problems in cosmology is why there is an
accelerated expansion of the universe.  Usually the explanation
given is an exotic energy known as dark energy.  On the other hand,
instead of looking for candidates for such energy, alternative
explanations can be tried.  This last chain of thought will be used. The more interesting features of the universe are analyzed.
There is a non-zero temperature other than zero even at the
most distant point in interstellar space.  In addition, it has an
observable dynamic horizon increasing with time.  Such a horizon
works as a causal barrier to events within it.  Mainly this system behaves like a spherical Casimir effect.  There appears to be two associated phenomena observed in
the universe: i) a thermal radiation like Stefan-Boltzmann's law and
ii) a force \textit{a la} Casimir effect responsible for an
accelerated expansion of the system.  This leads us to consider that
gravitation at finite temperature explains such a phenomena.  This hypothesis is explored here.

This article is divided as follows.  In section \ref{sec.2} TFD is introduced briefly.  In section \ref{sec.3} the TEGR is presented
and the thermal expressions are calculated.  In section \ref{sec.4}
the energy-momentum tensor at finite temperature is applied to the FRW
universe.  With this both the Stefan-Boltzmann law and the Casimir effect
for a zero curvature in the metric are calculated.  Finally conclusions are presented in the last section.

\section{Thermo Field Dynamics (TFD)} \label{sec.2}

A quantum field theory at finite temperature is developed by two
distinct, but equivalent, approaches: (i) the imaginary time
formalism \cite{matsubara} and (ii) the real time formalism
\cite{Schwinger, Umezawa1, Umezawa2, Umezawa22, Khanna1, Khanna2}.
TFD is a real-time finite temperature formalism. The temperature dependent vacuum is defined such that the vacuum
expectation value of an arbitrary operator $A$ agrees with the
statistical average, i.e., \bea \langle A \rangle=\langle 0(\beta)|
A|0(\beta) \rangle, \eea where  $|0(\beta) \rangle$ is the thermal
vacuum and $\beta=\frac{1}{k_BT}$, with $T$ being the temperature
and $k_B$ the Boltzmann constant. To construct this thermal state
two elements are necessary: the doubling of the original Hilbert
space and the Bogoliubov transformation. This doubling is defined by
${\cal S}_T={\cal S}\otimes \tilde{\cal S}$, where ${\cal S}$ is the
Hilbert space and $\tilde{\cal S}$ is the dual (tilde) space. The
map between the non-tilde ${A_i}$ and tilde $\tilde{A_i}$ operators
is given by tilde (or dual) conjugation rules. These rules are \bea
(A_iA_j)^\thicksim &=&
\tilde{A_i}\tilde{A_j},\nonumber\\(cA_i+A_j)^\thicksim &=&
c^*\tilde{A_i}+\tilde{A_j}, \nonumber\\ (A_i^\dagger)^\thicksim &=&
\tilde{A_i}^\dagger, \nonumber\\ (\tilde{A_i})^\thicksim &=& -\xi
A_i, \eea with $\xi = -1 (+1)$ for bosons (fermions). In addition,
the tilde conjugation rules associate each operator in ${\cal S}$ to
two operators in ${\cal S}_T$. Considering $a$ as an operator leads to
\bea
 A=a\otimes 1,\quad\quad\quad\quad \tilde{A}=1\otimes a.
\eea

TFD and Bogoliubov transformations introduce thermal effects through a rotation
between tilde ($\tilde{\cal S}$) and non-tilde (${\cal S}$)
operators. With an arbitrary operator ${\cal O}$, the Bogoliubov
transformation is defined as \bea \left( \begin{array}{cc} {\cal
O}(k, \alpha)  \\\xi \tilde {\cal O}^\dagger(k, \alpha) \end{array}
\right)={\cal B}(\alpha)\left( \begin{array}{cc} {\cal O}(k)  \\
\xi\tilde {\cal O}^\dagger(k) \end{array} \right), \eea where the
$\alpha$ is called the compactification parameter defined
by $\alpha=(\alpha_0,\alpha_1,\cdots\alpha_{D-1})$ and ${\cal
B}(\alpha)$ is \bea
{\cal B}(\alpha)=\left( \begin{array}{cc} u(\alpha) & -w(\alpha) \\
\xi w(\alpha) & u(\alpha) \end{array} \right), \eea with
$u^2(\alpha)+\xi w^2(\alpha)=1$. These quantities $u(\alpha)$ and
$w(\alpha)$ are related to the Bose distribution. For the case
$\alpha_0\equiv\beta$ and $\alpha_1,\cdots\alpha_{D-1}=0$, the
temperature effect is introduced. Using such formalism, a
topological quantum field theory is considered. A topology
$\Gamma_D^d=(\mathbb{S}^1)^d\times \mathbb{R}^{D-d}$ with $1\leq d
\leq D$ is used. Here $D$ is the space-time dimensions
and $d$ is the number of compactified dimensions. Any set of
dimensions of the manifold $\mathbb{R}^{D}$ can be compactified.

In the TFD formalism, all propagators are written in terms of the
compactification parameter  $\alpha $. Here the
scalar field propagator is defined as \bea
G_0^{(AB)}(x-x';\alpha)=i\langle 0,\tilde{0}|
\tau[\phi^A(x;\alpha)\phi^B(x';\alpha)]| 0,\tilde{0}\rangle, \eea
where $\tau$ is the time ordering operator and $A\, \mathrm{and}
\,B=1,2$. The Bogoliubov transformation
is used to write as \bea \phi(x;\alpha)&=&{\cal B}(\alpha)\phi(x){\cal
B}^{-1}(\alpha). \eea In the thermal vacuum, which is
defined as $|0(\alpha)\rangle={\cal U}(\alpha)|0,\tilde{0}\rangle$,
the propagator becomes \bea
G_0^{(AB)}(x-x';\alpha)&=&i\langle 0(\alpha)| \tau[\phi^A(x)\phi^B(x')]| 0(\alpha)\rangle,\nonumber\\
&=&i\int \frac{d^4k}{(2\pi)^4}e^{-ik(x-x')}G_0^{(AB)}(k;\alpha),
\eea where \bea G_0^{(AB)}(k;\alpha)={\cal
B}^{-1}(\alpha)G_0^{(AB)}(k){\cal B}(\alpha), \eea with \bea
G_0^{(AB)}(k)=\left( \begin{array}{cc} G_0(k) & 0 \\
0 & \xi G^*_0(k) \end{array} \right), \eea and \bea
G_0(k)=\frac{1}{k^2-m^2+i\epsilon}, \eea where $m$ is the mass. The Green function becomes \bea
G_0^{(11)}(k;\alpha)=G_0(k)+\xi w^2(k;\alpha)[G^*_0(k)-G_0(k)]. \eea
Here the physical quantities are given by the non-tilde
variables, i.e. $A=B=1$. In addition, $w^2(k;\alpha)$ is the
generalized Bogoliubov transformation \cite{GBT} given as
\bea
w^2(k;\alpha)=\sum_{s=1}^d\sum_{\lbrace\sigma_s\rbrace}2^{s-1}\sum_{l_{\sigma_1},...,l_{\sigma_s}=1}^\infty(-\xi)^{s+\sum_{r=1}^sl_{\sigma_r}}\,\exp\left[{-\sum_{j=1}^s\alpha_{\sigma_j}
l_{\sigma_j} k^{\sigma_j}}\right],\label{BT} \eea where
$\lbrace\sigma_s\rbrace$ denotes the set of all combinations with
$s$ elements and $k$ is the 4-momentum.

\section{Teleparallel Gravity} \label{sec.3}

Teleparallelism Equivalent to General Relativity (TEGR) is dynamically equivalent to the standard theory of
gravitation formulated in a Riemann space.  However, TEGR is
described in terms of torsion in the Weitzenb\"ock space. The
connection in such a space is
$$\Gamma_{\mu\lambda\nu}=e^{a}\,_{\mu}\partial_{\lambda}e_{a\nu}\,,$$
where $e^{a}\,_{\mu}$ is the tetrad field. It is the dynamical
variable of the theory. The relationship between the metric tensor
and the tetrad field is $g_{\mu\nu}=e^{a}\,_\mu
e_{a\nu}\,.$ The tetrad contains two important symmetries, that is the bridge between them.  Lorentz symmetry (Latin indices) and the transformation of coordinates (Greek indices). It is interesting to note that the
Weintzenb\"ock connection is curvature free, while the
anti-symmetric part establishes the following torsion tensor
\begin{equation}
T^{a}\,_{\lambda\nu}=\partial_{\lambda} e^{a}\,_{\nu}-\partial_{\nu}
e^{a}\,_{\lambda}\,. \label{4}
\end{equation}
This connection is related to the Christoffel symbols by
\begin{equation}
\Gamma_{\mu \lambda\nu}= {}^0\Gamma_{\mu \lambda\nu}+ K_{\mu
\lambda\nu}\,, \label{2}
\end{equation}
where the contortion tensor, $K_{\mu \lambda\nu}$, is given by
\begin{eqnarray}
K_{\mu\lambda\nu}&=&\frac{1}{2}(T_{\lambda\mu\nu}+T_{\nu\lambda\mu}+T_{\mu\lambda\nu})\,,\label{3}
\end{eqnarray}
with $T_{\mu \lambda\nu}=e_{a\mu}T^{a}\,_{\lambda\nu}$. The above identity leads to the relation
\begin{equation}
eR(e)\equiv
-e(\frac{1}{4}T^{abc}T_{abc}+\frac{1}{2}T^{abc}T_{bac}-T^aT_a)+2\partial_\mu(eT^\mu)\,.\label{eq5}
\end{equation}
The Lagrangian density for TEGR is
\begin{eqnarray}
\mathfrak{L}(e_{a\mu})&=& -\kappa\,e\,(\frac{1}{4}T^{abc}T_{abc}+
\frac{1}{2} T^{abc}T_{bac} -T^aT_a) -\mathfrak{L}_M\nonumber \\
&\equiv&-\kappa\,e \Sigma^{abc}T_{abc} -\mathfrak{L}_M\;, \label{6}
\end{eqnarray}
where $\kappa=1/(16 \pi)$, $\mathfrak{L}_M$ is the Lagrangian
density of matter fields and $\Sigma^{abc}$ is given by
\begin{equation}
\Sigma^{abc}=\frac{1}{4} (T^{abc}+T^{bac}-T^{cab}) +\frac{1}{2}(
\eta^{ac}T^b-\eta^{ab}T^c)\;, \label{7}
\end{equation}
with $T^a=e^a\,_\mu T^\mu$. If a derivative of eq. (\ref{6}) with
respect to the tetrad field is performed, the field equation
reads
\begin{equation}
\partial_\nu\left(e\Sigma^{a\lambda\nu}\right)=\frac{1}{4\kappa}
e\, e^a\,_\mu( t^{\lambda \mu} + T^{\lambda \mu})\;, \label{10}
\end{equation}
where
\begin{equation}
t^{\lambda \mu}=\kappa\left[4\,\Sigma^{bc\lambda}T_{bc}\,^\mu-
g^{\lambda \mu}\, \Sigma^{abc}T_{abc}\right]\,, \label{11}
\end{equation}
is the gravitational energy-momentum tensor. Such an expression is frame dependent and to calculate its average a class of observers must be chosen, that is, certain conditions must be imposed on the tetrad field . It is to be noted that the skew-symmetry in $\Sigma^{a\lambda\nu}$ leads to
\begin{equation}
\partial_\lambda\partial_\nu\left(e\Sigma^{a\lambda\nu}\right)\equiv0\,.\label{12}
\end{equation}
This is the conservation law. It is then possible to
establish the energy-momentum vector as
\begin{equation}
P^a = \int_V d^3x \,e\,e^a\,_\mu(t^{0\mu}+ T^{0\mu})\,, \label{14}
\end{equation}
or with the help of eq. (\ref{10}), it reads
\begin{equation}
P^a =4k\, \int_V d^3x
\,\partial_\nu\left(e\,\Sigma^{a0\nu}\right)\,. \label{14.1}
\end{equation}
This is the total energy vector. It is interesting to note
that it is a vector under global Lorentz transformation which implies that
energy, as the zero component of this 4-vector, is not an invariant. In fact, it depends on the choice of tetrad, which determines the very choice of the observer.
On the other hand the quantity is not dependent on the coordinate
choice. These are indeed desirable features for any definition of
energy-momentum.

With the well-established definition of an energy-moment tensor, the first element necessary for the application of TFD is defined.  It is
still necessary to obtain a propagator for the field.  Using the weak field approximation
\begin{equation}
g_{\mu\nu}=\eta_{\mu\nu}+h_{\mu\nu},
\end{equation}
which in eq. (\ref{6}) leads to
\begin{equation}
\langle e_{b\lambda}, e_{d\gamma} \rangle=\Delta_{bd\lambda\gamma} =
\frac{\eta_{bd}}{\kappa q^{\lambda} q^{\gamma}}.
\end{equation}
This is the graviton propagator \cite{usk}. Then the Green function is
 \bea
G_0(x,x')=-i\Delta_{bd\lambda\gamma}\,g^{\lambda\gamma}\eta^{bd}.
\eea Explicitly it is
\begin{equation}
G_0(x,x')= -\frac{i64\pi}{q^{2}}\,,
\end{equation}
with $q=x-x'$, where $x$ and $x'$ are four vectors. With the
weak field  approximation the gravitational energy-momentum tensor
$t^{\lambda \mu}$ becomes
\begin{eqnarray}
t^{\lambda\mu}(x) &=& \kappa\Bigl[g^{\mu\alpha}\partial^{\gamma}e^{b\lambda}\partial_{\gamma}e_{b\alpha} - g^{\mu\gamma}\partial^{\alpha}e^{b\lambda}\partial_{\gamma}e_{b\alpha} - g^{\mu\alpha}(\partial^{\lambda}e^{b\gamma}\partial_{\gamma}e_{b\alpha} - \partial^{\lambda}e^{b\gamma}\partial_{\alpha}e_{b\gamma})\nonumber\\
        & &-2g^{\lambda\mu}\partial^{\gamma}e^{b\alpha}(\partial_{\gamma}e_{b\alpha}-\partial_{\alpha}e_{b\gamma})\Bigl]\,.
\end{eqnarray}
For dealing with the mean of the
energy-moment tensor the standard procedure is to consider it at different points of the
space and then take the limit. This avoids
divergences. Hence \bea
\langle t^{\lambda\mu}(x)\rangle&=& \langle 0|t^{\lambda\mu}(x)|0\rangle,\nonumber\\
&=& \lim_{x^\mu\rightarrow x'^\mu}
4i\kappa\left(-5g^{\lambda\mu}\partial'^{\gamma}\partial_{\gamma}
+2g^{\mu\alpha}\partial'^{\lambda}\partial_{\alpha}\right)G_{0}(x-x')\,,\label{em}
\eea where $\langle e_{c}^{\,\,\,\lambda}(x), e_{b\alpha}(x')
\rangle = i\eta_{cb}\,\delta^{\lambda}_{\alpha}\,G_{0}(x-x')$.
This average applies to any metric that is related to the linearized
Einstein's equations.  On the other hand, the validity of this
expression is restricted to stationary observers.

\section{Stefan-Boltzmann law and Casimir effect in FRW universe } \label{sec.4}

The TEGR expression in the weak field
approximation leads to the TFD framework. The mean value of the
energy-moment tensor becomes
\bea \langle t^{\lambda\mu(AB)}(x;\alpha)\rangle=\lim_{x\rightarrow
x'}
4i\kappa\left(-5g^{\lambda\mu}\partial'^{\gamma}\partial_{\gamma}
+2g^{\mu\alpha}\partial'^{\lambda}\partial_{\alpha}\right)G_{0}^{(AB)}(x-x';\alpha).
\eea If we use the Casimir prescription, \bea {\cal T}^{\lambda\mu
(AB)}(x;\alpha)=\langle t^{\lambda\mu(AB)}(x;\alpha)\rangle-\langle
t^{\lambda\mu(AB)}(x)\rangle\,, \eea then \bea {\cal T}^{\lambda\mu
(AB)}(x;\alpha)=\lim_{x\rightarrow
x'}\Gamma^{\lambda\nu}(x,x')\overline{G}_{0}^{(AB)}(x-x';\alpha),\label{EM}
\eea 
where
\bea\label{gama}
\Gamma^{\lambda\nu}=4i\kappa\left(-5g^{\lambda\mu}\partial'^{\gamma}\partial_{\gamma}
+2g^{\mu\alpha}\partial'^{\lambda}\partial_{\alpha}\right)\,, \eea
and \bea
\overline{G}_0^{(AB)}(x-x';\alpha)=G_0^{(AB)}(x-x';\alpha)-G_0^{(AB)}(x-x')\,.
\eea It is necessary to establish the appropriate space-time
geometry i.e.,  analysing the result of such expressions on
cosmological scales.  A homogeneous and isotropic
universe is chosen. The suitable line element  is
\bea\label{FRW}
ds^{2}=-dt^{2}+a\left(t\right)\left(dr^{2}+r^{2}d\theta^{2}+r^{2}\sin^2{\theta}d\phi^{2}\right)\,,
\eea which is the FRW line element of zero curvature.  This metric
respects the approach used, as well as the constraints arising from the
experiments. If eq. (\ref{gama}) is used
together with eq. (\ref{FRW}), then
\bea
\Gamma^{00}=\frac{i}{4\pi}\left[-3\partial'_{0}\partial_0+\frac{5}{a^{2}}\left(\partial'_{1}\partial_1+\frac{1}{r^{2}}\partial'_{2}\partial_{2}+
\frac{1}{r^{2}\sin^2{\theta}}\partial'_{3}\partial_{3}
\right)\right] \eea and \bea \Gamma^{11} = \frac{i}{4\pi
a^2}\left[5\partial'_{0}\partial_0-\frac{5}{a^{2}}\left(\frac{3}{5}\partial'_{1}\partial_1+\frac{1}{r^{2}}\partial'_{2}\partial_{2}+
\frac{1}{r^{2}\sin^2{\theta}}\partial'_{3}\partial_{3}
\right)\right]\,. \eea Using these relations to calculate the
energy and pressure for Stefan-Boltzmann law and the
Casimir effect according to the Bogoliubov transformation is
desired.

\subsection{Gravitational Stefan-Boltzmann Law }

To calculate the Stefan-Boltzmann law, $\alpha=(\beta,0,0,0)$ is chosen, which leads to the Bogoliubov
transformation
\bea v^2(\beta)=\sum_{j_0=1}^\infty e^{-\beta k^0 j_0}\,, \eea where $\beta=\frac{1}{T}$. 
Then the Green function is
\bea \overline{G}_0^{(11)}(x-x';\beta)&=&2\sum_{j_0=1}^\infty
G_0^{(11)}\left(x- x'-i\beta j_0 n_0\right),\label{1GF} \eea where
$n_0=(1,0,0,0)$ and the physical component $(AB)=(11)$ is chosen,
then
\bea E=\frac{32\pi^{4}}{15}T^{4}\,, \eea and \bea
P=\frac{32\pi^{4}}{45a^{2}}T^{4}\,, \eea  with $E=\langle t^{00(11)}(x;\beta)
\rangle $ and $P=\langle t^{11(11)}(x;\beta) \rangle $.
It is interesting to note that the pressure is dependent on the
scale factor which in turn is expanded as
\bea
a=1+H_{0}\left(t-t_{0}\right)-\frac{q_0 H_{0}^{2}}{2}\left(t-t_{0}\right)^{2}\,,
\eea
where $H_0$ and $q_0$ refer to the Hubble constant and the
deceleration parameter respectively. So when $a = 1$,
the state equation becomes $P = \frac{E}{3}$, which is to be the expected
state equation for the graviton. Taking into account the
relation $\left(\frac{\partial P}{\partial
T}\right)_{V}=\left(\frac{\partial S}{\partial V}\right)_{T}$, the entropy density is
\bea s=\frac{S}{V}=\frac{128\pi^{4}}{45a^{2}}T^{3}\,. \eea Using the
expansion for the scale factor above, the second time derivative of the
entropy density is
\bea
\ddot{s}=-\frac{256\pi^{4}T^{3}}{45a^{2}}\left[\frac{\ddot{a}}{a}-3\left(\frac{\dot{a}}{a}\right)^{2}\right]\,,
\eea where dot means a time derivative. Here the Landau
theory of second order phase transition is involved.  A
divergence in the second derivative of the entropy determines a critical
quantity that characterizes the phase transition.  Here it is assumed that time is the dynamic variable.
Hence $s\rightarrow\infty$ implies $a=0$. If 
$\tau=H_{0}\left(t-t_{0}\right)$ is defined as an auxiliary variable, then it
follows that
\bea
\frac{q_0}{2}\tau^{2}-\tau-1=0\,.
\eea
This imposes a constraint on the current deceleration parameter,
such that $q_0\geq-\frac{1}{2}\,.$ This is an interesting result
considering that the deceleration parameter is written in terms
of the main cosmological parameters. Results obtained in
\cite{Feeney} showed that using the broad (truncated) Gaussian 
$q_0 = - 0.5 \pm 1$, it is indeed possible to obtain a competitive
constraint on the Hubble constant.  These results are consistent
with phenomenological models of the interaction rates \cite{Pan}
using the latest microwave background observations from Planck 2018
and baryon acoustic oscillations measurements.

\subsection{Casimir effect}

The Casimir effect is described in TFD with the choice $\alpha=(0,i2d,0,0)$, where $d$
is the radius of the outer spherical surface.
This leads to the Bogoliubov transformation

\bea v^2(d)=\sum_{l_1=1}^\infty e^{-i2dk^1l_1}. \eea If the Green
function is given by \bea
\overline{G}_0^{(11)}(x-x';d)&=&2\sum_{l_1=1}^\infty G_
0^{(11)}\left(x-x'-2dl_1n_1\right),\label{2GF} \eea with
$n_1=(0,1,0,0)$, then \bea E_c=-\frac{2\pi^{4}}{45d^4a^4}\,, \eea
and \bea P_c= -\frac{2\pi^{4}}{15d^4a^6}\,. \eea This result is
obtained by choosing the physical component of Green's function
$(AB)=(11)$. The same identification for the average energy-momentum
tensor, $E_c=\langle t^{00(11)}(x;d) \rangle $ and
$P_c=\langle t^{11(11)}(x;d) \rangle $. Two important features
need to be highlighted.  The first is the Casimir energy and
pressure are obtained in a vacuum.  A time-dependent negative
pressure is consistent with an accelerating expanding universe.  The
second is that Casimir pressure is associated with the
cosmological constant $\Lambda$.  In	 natural
units, the pressure and the cosmological constant have the same dimension. Thus the cosmological constant is
understood as a fluid with the following pressure
$$p=-\frac{c^4\Lambda}{G}\,,$$ thus this given by, in unities of the international system,

\bea \Lambda= \frac{2\pi^{4}G\hbar}{15d^4c^3}\,, \eea in the present
time. In this estimate the outer surface is used as the observable
radius of the universe, this determines $d$ as $10^{10}$ light
years.  The cosmological constant is of the order of $10^{-180}\,m^{-2}$. It is interesting to note that in an
incipient universe $\Lambda$ was much larger than it is today.

\section{Conclusion} \label{sec.5}

The Stefan-Boltzmann law and the Casimir effect are analyzed in a homogeneous and isotropic universe. The FRW metric for zero curvature is used. From the Stefan-Boltzmann law it is possible to understand that there is an energy and pressure from strictly gravitational thermal radiation. The entropy density provides for a phase transition that limits the range of the deceleration parameter. The Casimir effect establishes a negative pressure consistent  with an accelerated expanding universe. Such a quantity when interpreted as the observable radius of the universe leads to the conclusion that the cosmological constant is  small. It is important to note that due to its temporal evolution, the cosmological constant played a more relevant role in a primordial universe. When Casimir effect is established at finite temperature, imaginary quantities are obtained. This leads to interpret that the temperature effect in the universe is independent of the pressure exerted by the vacuum. Perhaps both effects are linked on a smaller scale when quantum effects are more relevant.

\section*{Acknowledgments}

This work by A. F. S. is supported by CNPq projects 308611/2017-9 and 430194/2018-8.


\begin{thebibliography}{99}

\bibitem{gravTFD}
S. C. Ulhoa, A. F. Santos, T. F. Furtado, F. C. Khanna, Advances in
High Energy Physics, {\bf 2019}, p. 1-6, (2019).

\bibitem{matsubara}
 T. Matsubara, Prog. Theor. Phys. {\bf 14}, 351 (1955).

\bibitem{flatTFD}

S. C. Ulhoa, A. F. Santos and F. C. Khanna, General Relativity and
Gravitation {\bf 49},  54, (2017).

\bibitem{4}
G. Acquaviva, A. Iorio, M. Scholtz. Annals of Physics, {\bf 387}, 317-333, (2017).

\bibitem{5}
G. Acquaviva, A. Iorio and L. Smaldone, Physical Review D, {\bf 102}, 106002, (2020).


\bibitem{maluf}
J. W. Maluf, Annalen der Physik, {\bf 525}, no. 5, pp. 339-357,
(2013).

\bibitem{entropy}
S. C. Ulhoa, E. P. Spaniol, R. Gomes,  A. F. Santos, A. E. Santana,
Advances in High Energy Physics, {\bf2020}, p. 1-9, (2020).

\bibitem{7}
M. Krssak, R. J. van den Hoogen, J. G. Pereira, C. G. B\"{o}hmer and A. A. Coley, Class. Quant. Grav. {\bf 36}, no.18, 183001, (2019).

\bibitem{8}
E. D. Emtsova, A. N. Petrov and A. V. Toporensky, Class. Quant. Grav. {\bf 37}, no.9, 095006, (2020).

\bibitem{9}
M. Hohmann, L. Jarv, M. Krssak and C. Pfeifer, Phys. Rev. D {\bf 100}, no.8, 084002, (2019).

\bibitem{10}
 J W Maluf, S C Ulhoa, J F da Rocha-Neto and F L Carneiro, Class. Quantum Grav. {\bf 37} 067003 (2020).

\bibitem{Schwinger}J. Schwinger, J. Math. Phys. {\bf 2}, 407 (1961); J. Schwinger, Lecture Notes Of Brandeis University
Summer Institute (1960).
\bibitem{Umezawa1}Y. Takahashi and H. Umezawa, Coll. Phenomena {\bf 2}, 55 (1975); Int. Jour. Mod. Phys. B {\bf 10}, 1755 (1996).
\bibitem{Umezawa2}Y. Takahashi, H. Umezawa and H. Matsumoto, Thermofield Dynamics and Condensed States, North-Holland, Amsterdan, (1982); F. C. Khanna, A. P. C. Malbouisson, J. M. C. Malboiusson and A. E. Santana, Themal quantum field theory: Algebraic aspects and applications, World Scientific, Singapore, (2009).
\bibitem{Umezawa22} H. Umezawa, Advanced Field Theory: Micro, Macro and Thermal Physics, AIP, New York, (1993).
\bibitem{Khanna1} A. E. Santana and F. C. Khanna, Phys. Lett. A {\bf 203}, 68 (1995).
\bibitem{Khanna2} A. E. Santana, F. C. Khanna, H. Chu, and C. Chang, Ann. Phys. {\bf 249}, 481 (1996).
\bibitem{GBT}F. C. Khanna, A. P. C Malbouisson, J. M. C. Malbouisson and A. E. Santana, Ann. Phys. {\bf 326}, 2634 (2011).
\bibitem{usk}
S. C. Ulhoa, A. F. Santos and F. C. Khanna. International Journal of Theoretical Physics, p. 1995, (2017).

\bibitem{Feeney} S. M. Feeney, D. J. Mortlock, and N. Dalmasso, Mon. Not. R. Astron. Soc. {\bf 476}, 3861 (2018).
\bibitem{Pan} S. Pan, G. S. Sharov, and W. Yang, Phys. Rev. D {\bf 101}, 103533 (2020)

\end{thebibliography}
\end{document}